\documentclass[twocolumn,prl]{revtex4}
\usepackage{psfig}
\usepackage{epsfig}
\usepackage{amsmath}
\usepackage{graphicx}

\input{tcilatex}

\begin{document}

\title{Interdependence of dynamical signals and topology: Detecting the
influential nodes in networks}
\author{Lei Yang$^{2,3,4}$, Liang Huang$^{2}$, Yong Zhang$^{2,3,4}$ and
Kongqing Yang$^{1,2}$}
\affiliation{$^{1}$Institute of Applied Physics, Jimei University, Xiamen 361021, China}
\affiliation{$^{2}$Department of Physics, Lanzhou University, Lanzhou 730000, China}
\affiliation{$^{3}$Centre for Nonlinear Studies, Hong Kong Baptist University, Hong Kong,
China}
\affiliation{$^{4}$The Beijing-Hong Kong-Singapore Joint Centre for Nonlinear and Complex
Systems (HK Branch), Hong Kong Baptist University, Hong Kong, China}

\begin{abstract}
By studying varies dynamical processes, including coupled maps, cellular
automata and coupled differential equations, on five different kinds of
known networks, we found a positive relation between signal correlation and
node's degree. Thus a method of identifying influential nodes in dynamical
systems is proposed, its validity is studied, and potential applications on
real systems are discussed.
\end{abstract}

\date{\today}
\maketitle

Dynamical processes on networks have been highly concerned \cite%
{ws,watts0,strogatz,i}, the network topological structure often plays a
crucial role in determining the system's dynamical features. Various
dynamical systems on complex networks have been widely investigated, such as
transient activation in neuron networks \cite{lago}, synchronization \cite%
{wang,barahona}, virus spreading or information diffusion \cite{virus}. Due
to the complex connectivity and the dynamical nonlinearity, an exact
mathematical analysis of general dynamical processes on networks is often
highly intricate. In general dynamical systems on complex networks, the
collective behavior of nodes remains unclear.

Recently, studying dynamical signals (the time series) presents a different
way to understand the dynamical processes on complex networks. For example,
Ref. \cite{fluc} shows effectivity to simultaneously characterize the
dynamics of several thousands nodes by scaling of fluctuations; Ref. \cite%
{recon} gives an example to reconstruct the nonlinear dynamics in a network
from observed dynamical signals. Beside theoretical interests, studying the
dynamical signal may provide practical applications, i.e., for some
dynamical systems, it would be difficult or even impossible to directly
obtain the topological information of the background networks, such as
financial systems \cite{zhang}, the neuronal functional networks \cite%
{neuronal}, etc., thus a key problem arises that how to obtain the
topological information from the dynamical signals.

In the paper, based on principal component analysis \cite{pca}, we have
investigated the cross-correlation matrix of the dynamical signals, which
reflects the interdependence of the nodes on the network, and studied the
relation between the cross-correlation matrix and the topological
connectivity. We found that in most cases there is a positive relation
between the nodes' degrees and the components of the principal eigenvector,
the eigenvector corresponded with the largest eigenvalue, of the
cross-correlation matrix, i.e., the larger degree often corresponds to
larger components. Thus, we suggest a method to obtain the topology from the
dynamical signals. Three types of coupled dynamical models, namely, coupled
maps, cellular automata and coupled differential equations, and five kinds
of complex networks are studied. To be specific, the dynamical models are
coupled logistic maps, avalanche processes \cite{bkw,sdp}, and
integrate-and-fire models \cite{if2}. The complex network models include the
Erd\H{o}s-R\'{e}nyi (ER) random graph \cite{er} (which has a Poisson degree
distribution), the Barab\'{a}si-Albert (BA) model \cite{ba} (which has a
power law degree distribution with exponent $\lambda =3.0$), the generalized
random graphs with scale-free (SF) degree distribution \cite{sfr}, the
lattice embedded SF model \cite{lesf}, and the nearest neighbor growth model 
\cite{nng} (which has an exponential degree distribution). In the
simulation, the exponent of degree distribution for SF random graph and
lattice embedded SF networks is fixed at $\lambda =3.0$ to make the results
comparable with that of BA model.

We start with a general expression of dynamics on a network. The connecting
network is described by its adjacent matrix $A$: $A_{ij}$ equals to $1$ if
node $i$ and node $j$ have a common edge (then $j$ is said to be a neighbor
of $i$ and vice versa) and $0$ if not, and $k_{i}=\sum_{j=1}^{N}A_{ij}$ is
the degree of node $i$, $x_{i}(t)$ is the output signal of node $i$. The
cross-correlation matrix is determined as $C_{ij}=\frac{\left\langle
x_{i}x_{j}\right\rangle -\left\langle x_{i}\right\rangle \left\langle
x_{j}\right\rangle }{\sigma _{i}\sigma _{j}}$, where $\sigma _{i}$ is the
variance: $\sigma _{i}=\sqrt{\left\langle x_{i}^{2}\right\rangle
-\left\langle x_{i}\right\rangle ^{2}}$, and the average is over a period of
length $L_{d}$. Here we concern only about the strength of the correlation
between the nodes, rather than the signs, thus we define the correlation
strength matrix: $\widetilde{C}_{ij}=\left\vert C_{ij}\right\vert $. Then
its principal eigenvector could be determined, and the relation between the
components of the principal eigenvector\ and the nodes' degrees is
investigated\ in detail.

\begin{figure}[th]
\begin{center}
\epsfig{file=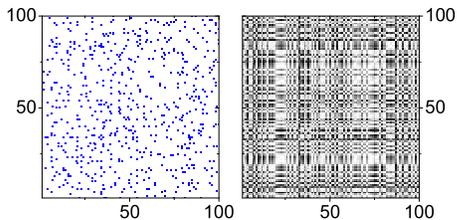, width=7cm}
\end{center}
\caption{(Color online) A direct show of network connection and correlation
strength of coupled maps on a ER random graph, with $N=100$, $\left\langle
k\right\rangle =6$. Left: Network connection, dot at ($i,j$) indicates the
connection of node $i$ and $j$. Right: Correlation strength, the brighter,
the stronger. $a=1.9$, $\protect\varepsilon =0.9$.}
\label{zzm}
\end{figure}

Coupled logistic maps. The dynamic of an individual node $i$ coupled with
its neighbors is described by: 
\begin{equation}
x_{i}^{t+1}=(1-\varepsilon )f(x_{i}^{t})+\frac{\varepsilon }{k_{i}}%
\sum_{j=1}^{N}A_{ij}f(x_{j}^{t}),  \label{cm}
\end{equation}%
where $f(x)$ is the logistic map $f(x)=1-ax^{2}$, $\varepsilon \in \left[ 0,1%
\right] $ is the coupling strength. For a given network, the parameter space
($\varepsilon ,a$) has two extreme cases: the synchronization regime and
complete non-synchronization regime. In the synchronization region, parts of
or all the nodes have the consistent motion. The cross-correlation matrix
may reflect the synchronization motion. Here, we investigate the system
running in the state which is far from synchronization. Figure \ref{zzm} is
a typical show of the connection matrix and the correlation strength matrix,
since the states have no particular orderliness, the latter is highly
noised. The same holds for other network models and for other dynamical
processes.

\begin{figure}[th]
\begin{center}
\epsfig{file=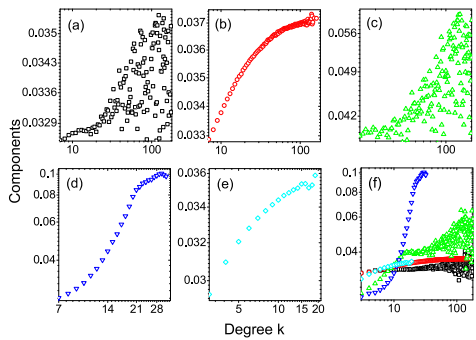, width=7cm}
\end{center}
\caption{(Color online) Components of principal eigenvector of correlation
strength matrices of coupled maps on networks vs. node degree for various
network models. (a) SF random graph; (b) BA model; (c) lattice embedded SF
model; (d) Nearest neighbor growth model; (e) ER random graph; (f) A
comparison of all the network models, same symbols represent the same
networks in (a)-(e).}
\label{ev1zz}
\end{figure}

\begin{figure}[th]
\begin{center}
\epsfig{file=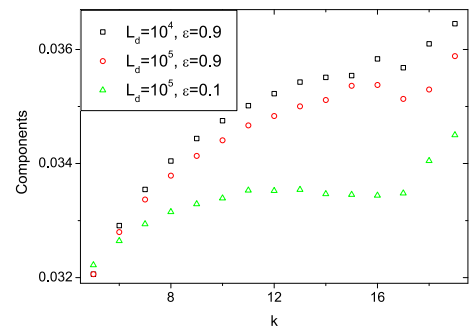, width=7cm}
\end{center}
\caption{(Color online) Components of principal eigenvector of correlation
strength matrices of coupled maps on networks vs. node degree for ER random
graphs, of different data length and different coupling strength. The
network size is $1000$, $<k>=6$, and $a=1.9$. The data has been averaged on $%
100$ ensembles. }
\label{ev1szz2}
\end{figure}

The relationship of the components of the principal eigenvector of the
correlation strength matrix and the nodes' degrees is shown in Fig. \ref%
{ev1zz}. In the simulation, the degree distribution exponent for SF networks
is $\lambda =3.0$, $<k>=6$ and the network size is $1000$. The dynamical
parameters are $a=1.9$, $\varepsilon =0.9$. The length of data used in
calculating correlation is $L_{d}=10^{5}$, and each group of data has been
averaged over $300$ network ensembles. The calculation of correlation begins
after $10000$ time steps, while the transition usually takes a few hundred
time steps. It should be noted that in the subgraphs the ranges of $x$-axes
are not the same due to the different degree distributions (SF, exponential,
and Poissonian). Although there are small deviations and the data scatter
sparsely for SF random graphs and lattice embedded\ SF networks, the
positive relation between the components and the degrees is clear. The
parameters in our study are chosen to satisfy that the dynamics are chaotic,
and the length of the data for calculating the correlation is long enough to
eliminate some unpredictable factors. In order to achieve this, we have
studied coupled maps on ER random graphs in detail, compared varies data
length and coupling strength. As Fig. \ref{ev1szz2} shows, the parameters
used in our study is appropriate. Similar studies on other network models
confirm that the relation is stable under variations of dynamical parameters.

The Bak-Tang-Wiesenfeld (BTW) sandpile model \cite{bkw} on SF networks has
been investigated by K. S. Goh \textit{et al.} \cite{sdp}, which set the
threshold of being toppled of each node $i$ be its degree $k_{i}$, unlike
that of the uniform threshold height in lattice cases; and proposed a losing
probability $f$ when the grains are transferring from a toppled node to its
neighbors, as the sinks or open boundaries in lattice cases. The rule can be
adopted on arbitrary networks. The correlation in this case is between the
toppling events $s_{i}^{n}$ in avalanches, $s_{i}^{n}=1$ if node $i$ toppled
in the $n$th avalanche, and $0$ if not. It is expected that the cooccurrence
between nodes in an avalanche could unveil the linkages between them.

\begin{figure}[th]
\begin{center}
\epsfig{file=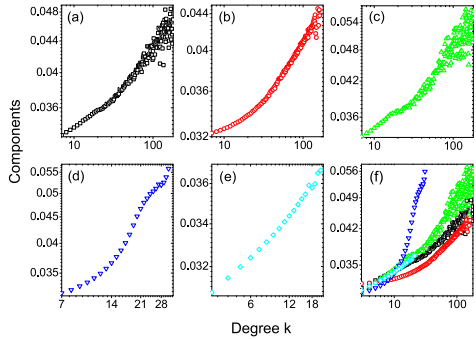, width=7cm}
\end{center}
\caption{(Color online) Components of principal eigenvector of correlation
strength matrices of the cooccurrence in avalanche events on networks vs.
node degree for various network models. The same panel indicates the same
network model as that in Fig. 2. }
\label{ev1sdp}
\end{figure}

The relationship of the components of principal eigenvector of the
correlation strength matrix of the cooccurrence in avalanche events on
networks and the nodes' degrees is shown in Fig. \ref{ev1sdp}. The network
parameters, such as the exponent of SF networks, average degree, and network
size, are the same with that in the coupled chaotic map cases. The loosing
probability is $f=0.001$. Again the length of the data is $L_{d}=10^{5}$,
and the correlation begins after $10000$ avalanche events, which is long
enough to avoid the transitions. Each data has been averaged over $300$
network ensembles. Also, the positive relation between component and degree
is clear.

\begin{figure}[th]
\begin{center}
\epsfig{file=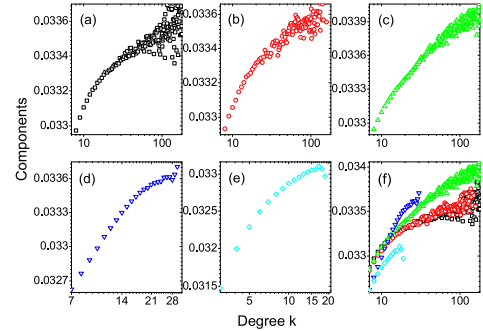, width=7cm}
\end{center}
\caption{(Color online) Components of the principal eigenvector of
correlation strength matrices of the currents for IFN model on networks vs.
node degree for various network models. The same panel indicates the same
network model as that in Fig. 2. }
\label{ev1slifi}
\end{figure}

Coupled integrate-and-fire neuron (IFN) model. The dynamics of each node is
described by \cite{if2} 
\begin{eqnarray*}
\frac{du_{i}(t)}{dt} &=&-u_{i}(t)+I_{0}+I_{i}(t), \\
\frac{dI_{i}(t)}{dt} &=&-\frac{I_{i}(t)}{\tau }+\varepsilon \frac{1}{k_{i}}%
\underset{j}{\tsum }A_{ij}\delta (t-t^{j}),
\end{eqnarray*}%
where $I_{0}$ is an external, time-independent and universal driven current, 
$\varepsilon $ is the coupling strength, and is positive in our simulation, $%
A$ is the adjacent matrix of the background network, as defined above, and $%
t^{j}$ is the spiking time of node $j$. When the potential $u_{i}(t)$
arrives at a threshold $u_{i}(t)=1$, it fires and resets to $0$. The IFNs
can be synchronized under certain conditions, which tells nothing about the
network topology. So we will focus on the unsynchronized states, and discuss
the current correlation, that is, the output signal of node $i$ is just its
current $I_{i}(t)$. The same relation is shown in Fig. \ref{ev1slifi}. The
parameters are: $I_{0}=1.3$, $\tau =0.5$, $\varepsilon =0.7$. The
correlation begins after a transient time of $T=100$ time units, and
performs for another $100$ time units. Beside the dispersion of some data,
the positive relation is apparent.

\begin{figure}[th]
\begin{center}
\epsfig{file=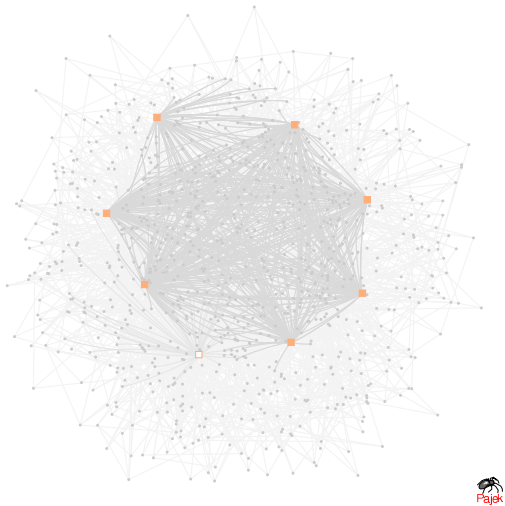, width=7cm}
\end{center}
\caption{(Color online) A direct show of the efficiency of our method.
Squares: the first $8$ nodes with largest degrees; filled squares: nodes
that are correctly located out by our method; empty square: the missed node,
which is just the $8$'th node among the first $8$ nodes; small circles:
ordinary nodes.}
\label{chn01n}
\end{figure}

We suggest a method to locate the important or influential nodes (namely,
the nodes that have large degrees) in dynamical systems on complex networks,
according to the positive relation between the components of the principal
eigenvector of the correlation strength matrix and the nodes' degrees.
First, one can measure the cross-correlation matrix from the output signal
of each node, and the component values of the principal eigenvector of the
correlation strength matrix could be calculated. The degree of a node is
represented by its corresponding component value. Thus, the important or
influential nodes are located as the nodes with larger component values. In
Fig. \ref{chn01n}, a typical result of the method is shown for coupled
chaotic maps on a BA network. For other dynamical and network models,
similar results can be obtained.

\begin{figure}[th]
\begin{center}
\epsfig{file=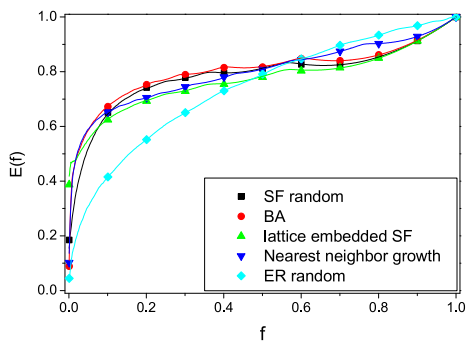, width=7cm}
\end{center}
\caption{(Color online) Efficiency of the locating method by ordering of the
component values of the principal eigenvector. The parameters are the same
as that of Fig. \protect\ref{ev1slifi}}
\label{match}
\end{figure}

Further more, we define the efficiency of the locating method as $%
E(f)=n_{m}(f)/(N\cdot f)$, where $N$ is the total number of the nodes, and $%
f $ is the fraction of selected nodes with largest degrees, and the $%
n_{m}(f) $\ is number of nodes that correctly matched. If the nodes were
randomly ordered, the fraction that matches with other ordering, say, by
degree, will be equal to the fraction that compared, thus $E(f)=f$. A
typical result by our method for IFNs is shown in Fig. \ref{match}. The
efficiency of the method boosts up rapidly as $f$ increases from $0$, and
remains high for almost all the region $\left[ 0,1\right] $. The high
efficiency of the method in IFNs model indicates direct application
potentials, i.e., identifying the most influencing neurons in the
experiments in neuronal networks \cite{neuronal}, (not necessarily through
current correlation).

In conclusion, we studied coupled chaotic maps, avalanche processes, and
integrate and fire neuron models, covering three types of coupled dynamical
systems, on various network topologies, whose degree distribution varies
from Poissonian to scale-free and to exponential, and of different local
properties. All our results support the proposal that the components of the
principal eigenvector of the correlation strength matrix have a positive
relation with the nodes' degrees, thus a representation of nodes' degrees,
the ability to influence others and the importance to the system, could be
realized by the components, which can be obtained through the dynamical
processes. Further applications such as immunization of internet or
contagious disease, identifying pivotal neurons, etc., can be investigated.

The work is supported by China National Natural Sciences Foundation with
grant 49894190 of major project and Chinese Academy of Science with grant
KZCX1-sw-18 of major project of knowledge innovation engineering. L. Yang
thanks the support of the Hong Kong Research Grants Council (RGC) and the
Hong Kong Baptist University Faculty Research Grant (FRG). K. Yang thanks
the support of Institute of Geology and Geophysics, CAS.

\end{document}